\documentstyle[12pt,epsfig]{article}
\def\beq{\begin{equation}}
\def\eeq{\end{equation}}
\def\bea{\begin{eqnarray}}
\def\eea{\end{eqnarray}}

\def\del{\partial}

\def\a{\alpha}

\def\beq{\begin{equation}}
\def\eeq{\end{equation}}
\def\ber{\begin{eqnarray}}
\def\eer{\end{eqnarray}}
\def\bea{\begin{eqnarray}}
\def\eea{\end{eqnarray}}
\def\del{\partial}

\def\a{\alpha}

\def\del{{\partial}}

\newfont{\Bbb}{msbm10 scaled 1200}     
\newcommand{\mathbb}[1]{\mbox{\Bbb #1}}

\def\lbldef#1#2{\expandafter\gdef\csname #1\endcsname {#2}}

\def\href#1#2{#2}

\newcommand{\beqar}{\begin{eqnarray}}

\newcommand{\eeqar}{\end{eqnarray}}

\def\del{\partial}

\def\a{\alpha}

\begin{document}

\pagestyle{plain} \setcounter{page}{1}
\begin{titlepage}

 \rightline{\small{\tt CALT-68-2377}}
 \rightline{\small{\tt CITUSC/02-010}}
 \rightline{\tt hep-th/0203257}

\vskip -1.0cm


\begin{center}

\vskip 2 cm

{\LARGE {Open Strings in PP-Wave Background \\
\vskip .2cm from Defect Conformal Field Theory}}

\vskip 2cm {\large Peter Lee and Jongwon Park}

\vskip 1.2cm

\end{center}

\begin{center}
\emph{ California Institute of Technology 452-48, Pasadena, CA
91125}

\vskip .7cm {\tt peter, jongwon@theory.caltech.edu}

\end{center}
\vspace*{1in}
\begin{center}
\textbf{Abstract}
\end{center}
\begin{quotation}
\noindent We consider open strings ending on a D5-brane in the
pp-wave background, which is realized in the Penrose limit of
$AdS_5 \times S^5$ with an $AdS_4\times S^2$ brane.  A complete
list of gauge invariant operators in the defect conformal field
theory is constructed which is dual to the open string states.
\end{quotation}
\vfil
\end{titlepage}

\newpage

\section{Introduction}
 AdS/CFT correspondence \cite{maldacena, gubser, witten} (for a review,
see \cite{ooguri2}) has lead to deep understandings of string
theory and field theory. However, until recently, much of the
progress in this direction has been limited to supergravity
approximations due to the difficulty when one has Ramond-Ramond
background.  Recently, it has been shown that string theory can be
fully solved in the pp-wave background even in the presence of RR
flux \cite{metsaev, metsaev2} in the light-cone Green-Schwarz
formalism. Shortly after this development, Berenstein, et. al.
\cite{BMN} have put forward an exciting proposal that tests
AdS/CFT correspondence beyond the supergravity approximation. More
specifically, they have related closed string states in the
pp-wave background with operators of the dual ${\cal N}=4$ SYM
with large R-charge $J \sim \sqrt{N}$ and finite $\Delta-J$.  Many
interesting papers have subsequently followed \cite{itzhaki,
ooguri3, russo, zayas, alishahia, kim, takayanagi, floratos,
billo, cvetic1, gursoy, michelson, rey, chu, cvetic2, dabholkar,
berkovits}.  \\

In this article, we extend the results of \cite{BMN} to the case
of open strings ending on a D5-brane in the pp-wave background. We
consider a large number of D3-branes and a single D5-brane in the
near-horizon limit.  The resulting system is $AdS_5\times S^5$
with the D5-brane spanning an $AdS_4\times S^2$. Recently extending
the idea of \cite{randall,randall2}, De Wolfe, et. al.
\cite{ooguri1} have proposed that its dual field theory is a
defect conformal field theory in which the usual ${\cal N}=4$ bulk
SYM theory is coupled to a 3-dimensional conformal defect. This
defect field theory has been further studied by \cite{erdmenger} in which they
demonstrate quantum conformal invariance for the non-abelian
case.  By taking the Penrose limit \cite{blau, penrose} of this
setup, one obtains a D5-brane in the pp-wave background. We
construct a complete list of gauge invariant operators in the
defect conformal
field theory which is dual to the open string states ending on the D5-brane.
Interestingly, boundary conditions of open stings on the D5-brane
are encoded in the way symmetry is broken by the defect and in specific
form of defect couplings in the dual field theory.\\

This paper is organized as follows.  In section 2, we give a brief
review of the D-brane setup and the field content of the defect
conformal field theory.  In section 3, we discuss the Penrose
limit of this background and obtain the open string spectrum.  In
the last section, we propose a list of gauge invariant operators
dual to the open string states. \\

While this manuscript was being prepared for publication, article
\cite{BMN2} containing some overlap with
our results was posted on the web.

\section{Review of defect conformal field theory}
In this section, we briefly review the D3-D5 brane setup of
\cite{randall,randall2} and the field content of its dual defect conformal
field theory discussed in \cite{ooguri1}. The interested reader
can find further details in the aforementioned papers. We start
with the coordinate system in which the world-volume of a stack of
$N$ D3-branes span the directions $(x^0,x^1,x^2,x^9)$ and a single
D5-brane spans the directions $(x^0,x^1,x^2,x^3,x^4,x^5)$. The
D-branes sit at the origin of their transverse coordinates. The
setup is summarized in the following table: \begin{center} \vskip
.1cm
\begin{tabular}{|c|c|c|c|c|c|c|c|c|c|c|} \hline
  &0&1&2&3&4&5&6&7&8&9 \\ \hline
D3&x&x&x& & & & & & &x \\ \hline D5&x&x&x&x&x&x& & & &  \\ \hline
\end{tabular} \vskip .2cm \end{center}
The presence of the D5-brane breaks 16 spacetime supersymmetries
to 8 supersymmetries and reduces the global symmetry group $SO(6)$
to $SO(3) \times SO(3)$, where each $SO(3)$ acts on the 345 and
678 coordinates respectively.  In AdS/CFT correspondence, one is
interested in taking the near horizon limit where the string
coupling $g \rightarrow 0$ and $N \rightarrow \infty$ with the
product $gN$ fixed. In this limit, we have the D5-brane spanning
an $AdS_4 \times S^2$ subspace of $AdS_5 \times S^5$.   The dual
conformal field theory of type IIB string theory in this
background is ${\cal N}=4$ SYM theory \cite{maldacena} that lives
on the boundary of $AdS_5$ parameterized by $(x^0,x^1,x^2,x^9)$.
The D5-brane introduces a codimension one conformal defect on this
boundary located at $x^9=0$.  An analogous model can be considered
for the $AdS_3\times S^3$ case, where an $AdS_2$ brane introduces
a one-dimensional defect in the dual CFT \cite{bachas}. Such a
reasoning has been used by \cite{lee, ponsot} to construct
boundary states for the $AdS_2$ branes.

It has been argued by DeWolfe, et. al. \cite{ooguri1} that type
IIB string theory in $AdS_5 \times S^5$ with a $AdS_4 \times S^2$
brane is dual to a {\it defect conformal field theory} wherein a
subset of fields of $d=4$, ${\cal N}=4$ SYM couples to a $d=3$,
${\cal N}=4$ $SU(N)$ fundamental hypermultiplet on the defect
preserving $SO(3,2)$ conformal invariance and 8 supercharges.  Let
us summarize the field content of the defect conformal field
theory relevant for our purposes. Denote the $SU(2)$ acting on the
345 directions as $SU(2)_H$ and the one acting on the 678
directions as $SU(2)_V$. Then we have the usual bulk $d=4$, ${\cal
N}=4$ vector multiplet which decomposes into a $d=3$, ${\cal N}=4$
vector multiplet and a $d=3$, ${\cal N}=4$ adjoint hypermultiplet.
The bosonic components of the vector multiplet are $A_\mu (\mu =
0,1,2) ,X^6,X^7,X^8$, with the scalars transforming as a {\bf 3}
of $SU(2)_V$, while those of hypermultiplet are $A_9,X^3,X^4,X^5$,
with the scalars as a {\bf 3} of $SU(2)_H$. The derivatives of
$X^3,X^4,X^5$ along the 9-direction, which is normal to the
defect, are also a part of the vector multiplet. The four adjoint
$d=4$ Majorana spinors of ${\cal N}=4$ SYM transform as a ({\bf
2},{\bf 2}) of $SU(2)_V \times SU(2)_H$, which is denoted as
$\lambda^{im}$. Under the dimensional reduction to  $d=3$, they
decompose into pairs of 2-component $d=3$ Majorana spinors,
$\lambda^{im}_1$ and $\lambda^{im}_2$, where the former is in the
vector multiplet and the latter in the hyper multiplet. We also
have a $d=3$, ${\cal N}=4$ $SU(N)$ fundamental hypermultiplet on
the defect. It consists of complex scalars $q^m$ transforming as a
{\bf 2} of $SU(2)_H$ and $d=3$ Dirac spinors $\Psi^i$ transforming
as a {\bf 2} of $SU(2)_V$. They are coupled canonically to
3-dimensional gauge fields $A_\mu$. Hence supersymmetry will
induce couplings to the rest of the bulk vector multiplet as well via defect
F-term,
while the bulk hypermultiplet does not couple to the defect
hypermultiplet at tree level. This fact will play a crucial role
in reproducing the open string spectrum in section 4. The field
content of interest is summarized in the following table.
\begin{center} \vskip .1cm
\begin{tabular}{|c|c|c|c|c|c|c|} \hline
   Field&Spin&$SU(2)_H$&$SU(2)_V$&$SU(N)$&$\Delta$ \\ \hline
   $X^3,X^4,X^5$&0&1&0&adjoint&1 \\ \hline
   $X^6,X^7,X^8$&0&0&1&adjoint&1 \\ \hline
   $\lambda^{im}$ & ${1\over2}$ & ${1\over2}$ & ${1\over2}$ & adjoint&
   ${3\over2}$ \\ \hline
   $q^m$&0&${1\over 2}$&0&$N$&${1\over 2}$ \\ \hline
   $\bar q^m$&0&${1\over 2}$&0&$\bar N$&${1\over 2}$ \\ \hline
   $\Psi^i$&${1\over 2}$&0&${1\over 2}$&$N$&$1$ \\ \hline
   $\bar\Psi^i$&${1\over 2}$&0&${1\over 2}$&$\bar N$&$1$ \\ \hline
\end{tabular}
\vskip .2cm \end{center} Field theory action takes the form
 \beq
  S = S_4 + S_3,
 \eeq
where $S_4$ is the usual $d=4$, ${\cal N}=4$ SYM part and $S_3$ is
the $d=3$ defect CFT action with defect couplings to $d=4$, ${\cal N}=4$ SYM
fields. They are derived in \cite{ooguri1} using the preserved $d=3$,
 ${\cal N}=4$ supersymmetry and the global symmetries. The
authors of \cite{ooguri1} convincingly argue that the chiral
primary operators in the defect CFT are
 \beq
    \bar q^m \tilde{\sigma}^{\left( I_0\right.}_{m n} X_H^{I_1}...X_H^{\left.I_J
  \right)}q^n,
 \eeq
where we define a {\it shifted} Pauli matrices $\tilde{\sigma}^I$
$(I=3,4,5)$ as $\sigma^{I-2}$ and $(...)$ denotes traceless
symmetrization. These operators will turn out to be the important
building blocks for open strings ending on the D5-brane in section
4.

\section{Open strings in pp-waves}
Let us now consider the Penrose limit of near-horizon geometry of
D3-D5 brane setup described in the previous section.  It is
convenient to introduce global coordinates on $AdS_5\times S^5$ in
taking the Penrose limit.  The metric takes the form
 \beq
  ds^2=R^2\left[-dt^2 \cosh^2\rho+d\rho^2+\sinh^2\rho d\Omega_3^2
  + d\psi^2 \cos^2\varphi + d\varphi^2 + \sin^2\varphi
  d\Omega^{'2}_3 \right],
 \eeq
where $R^4 = 4\pi g\alpha^{'2}N$. We introduce light-cone
coordinates $\tilde x^{\pm}=(t\pm \psi)/2$ and take the Penrose
limit ($R \rightarrow \infty$ with $g$ fixed ) after rescaling coordinates as
follows
 \beq
   \tilde{x}^+=x^+, \;\;\; \tilde x^-={x^-\over R^2}, \;\;\;
   \rho={r \over R}, \;\;\; \theta={y \over R}.
 \eeq
After rescaling $x^{\pm}$ to introduce a mass scale, $\mu$, the
metric and the Ramond-Ramond form takes the form
 \bea
   && ds^2 = -4 dx^+ dx^- - \mu^2 \vec{z}^{\;2}dx^{+2} + d\vec{z}^{\;2}, \\
   && F_{+1234}=F_{+5678}=\mu,
 \eea
where $\vec{z}=(z^1,...z^8)$.\footnote{We have chosen the null
geodesic in the Penrose limit to lie on the D5-brane because in
the light-cone gauge, Neumann conditions are automatically imposed
on $x^{\pm}$.} The SO(2) generator, $J=-i \del_\psi$, rotates the
34 plane in the original D3-D5 setup.  One finds that
 \bea
  2p^- &=& -p_+ = i\del_{x^+} = i\del_{\bar{x}^+} = i(\del_t
  +\del_\psi) = \Delta - J, \\
  2p^+ &=& -p_- = i\del_{x^-} = {i\over R^2}\del_{\bar{x}^-} = {i\over R^2}
  (\del_t-\del_\psi) = {\Delta + J \over R^2}.
 \eea
 Therefore, the Penrose limit corresponds to restricting to operators with
 large $J\sim \sqrt{N}$ and finite $\Delta -J$. Notice that we are in the large
't Hooft coupling regime since we keep $g$ fixed.

In the Penrose limit, the string action reduces to the following
form in the light-cone gauge
 \beq
  S = {1 \over 2\pi\a^\prime} \int d\tau \int_0 ^{\pi \a^\prime
  p^+} d\sigma \left[ {1\over2} \dot{z}^2 -{1\over2} z^{\prime 2}-
  {1\over2}\mu^2z^2 + i\left( {1\over 2} S_1 \del_+ S_1 + {1\over 2} S_2 \del_- S_2  - \mu
  S_1\Gamma^{1234}S_2\right) \right],
 \eeq
where $S_i$ are positive chirality $SO(8)$ spinors. One can
readily see that taking the light-cone gauge leads to Neumann
boundary conditions for $x^+,x^-$ in the open-string sector since
 \beq
  \del_\sigma x^-  = \frac{\del_\tau z^i \del_\sigma z^i}{p^+}.
 \eeq

We identify $(z^5,z^6,z^7,z^8)$ directions with the original
$(x^5,x^6,x^7,x^8)$ directions and $z^4$ with the orthogonal
direction to D5 brane in $AdS_5$. We label the coordinates in the
Penrose limit such that the boundary conditions for the D5-brane
are given as
\begin{center} \vskip .1cm
\begin{tabular}{|c|c|c|c|c|c|c|c|c|c|} \hline
+&-&1&2&3&4&5&6&7&8 \\ \hline N&N&N&N&N&D&N&D&D&D \\ \hline
\end{tabular}
\vskip .2cm \end{center} where N and D denote Neumann and
Dirichlet boundary conditions respectively. For $S_i$, the
appropriate boundary condition is \cite{west}
 \beq S_2 = \Gamma^{1235} S_1.
 \eeq
The boundary condition for the fermions effectively reduces the
degree of freedom by half.  Taking the Penrose limit and taking
the light-cone gauge break the symmetry group $SO(3,2)\times
SU(2)_H \times SU(2)_V$ to $SO(3)\times SU(2)_V$\footnote{This point has
been clarified in \cite{rey}.}.  The full open string spectrum on a
D5-brane has recently been computed by \cite{dabholkar}.  The mode
expansions for the bosonic part are
 \bea
  z^I_{NN}(\tau,\sigma) &=& \cos(\mu\tau) z^I_0 + {1\over
  \mu}\sin(\mu\tau)p^I_0
  +i\sum_{n=1}^\infty \frac{1}{\sqrt{\omega_n}} e^{-i\omega_n \tau}
  \cos\left(\frac{n\sigma}{\alpha^\prime p^+}\right) a^I_n + c.c., \\
  z^I_{DD}(\tau,\sigma) &=& i\sum_{n=1}^\infty \frac{1}{\sqrt{\omega_n}}
  e^{-i\omega_n \tau} \sin\left(\frac{n\sigma}{\alpha^\prime p^+}\right) a^I_n + c.c.,
 \eea
where we have defined
 \beq
  \omega_n = \sqrt{\mu^2+{n^2 \over 4(\a^\prime p^+)^2}}.
 \eeq
Important difference between the Neumann and Dirichlet expansions
is that the Dirichlet expansion does not have a zero mode.  This
gives rise to 4 massive bosonic oscillators.  Similarly, eight
zero modes coming from fermions form 4 massive fermionic
oscillators and their contribution to the zero point energy
exactly cancel the contribution from the bosonic zero modes. Due
to the mass term, fermionic creation and annihilation operators
have $+1/2$ and $-1/2$ eigenvalues with respect to $\Gamma^{45}$
respectively, and both transform separately as doublets of
$SU(2)_V$\footnote{This point is to be contrasted with \cite{BMN2}
where all creation operators have the same quantum number of the symmetry under
consideration.}. Hence, the light cone vacuum should be a singlet of
$SU(2)_V$ for the fermionic zero modes, thus correctly reproducing
D5-brane SYM multiplet.

The light cone Hamiltonian is given as
 \beq
   2p^- = -p_+ = H_{lc}= \sum_{n=0}^\infty N_n \sqrt{\mu^2+\frac{n^2}{4(\a^\prime
   p^+)^2}},
 \eeq
where $N_n$ denotes the total occupation number of that mode for
both bosonic and fermionic oscillators. Rewriting the Hamiltonian
in variables better suited for $AdS_5\times S^5$, one notes that a
typical string excitation contributes to $\Delta-J=2p^-$ with
frequency
 \beq
    \left(\Delta-J\right)_n = \sqrt{1+{\pi g N n^2 \over J^2}}.
 \eeq

\section{Open strings from defect conformal field theory}

In this section, we construct a list of gauge invariant operators
in the defect CFT dual to states in the open string Hilbert space.
Recall that $J$ is the generator of rotation on the $X^3$-$X^4$
plane.   Define
 \beq
  Z \equiv {1\over \sqrt{2}}\left(X^3+iX^4\right), \hskip 1cm  \sigma^Z
  \equiv {1\over \sqrt{2}} \left( \tilde{\sigma}^3 +i\tilde{\sigma}^4 \right)  =
  {1\over \sqrt{2}} \left( \sigma^1 + i\sigma^2 \right),
 \eeq
Both the operators $Z$ and $\bar{q}^m\sigma^Z_{mn}q^n$ have
$\Delta = J = 1$. The fact that $Z$ belongs to the bulk hypermultiplet will be important
later. We propose that the light-cone vacuum
corresponds to
 \beq \label{vacuum}
  |0,p^+\rangle_{l.c.}
  \longleftrightarrow \frac{1}{N^{J/2}} \sigma^Z_{mn} \bar{q}^m
  \underbrace{ZZ\cdots\cdots Z}_{J-1} q^n.
 \eeq
As mentioned above, this is a chiral primary operator with $\Delta
= J$ found in \cite{ooguri1}.  Because it is a chiral primary,
$\Delta - J = 0$ in the strong 't Hooft coupling limit. This
property agrees with the fact that the light-cone vacuum has zero
energy. Furthermore, it does not transform under $SU(2)_V$ as one
expects from the light-cone vacuum.

For excited states, as in the closed string case \cite{BMN}, we insert proper
operators with $\Delta - J = 1$ without phases for zero modes and
with appropriate phases for nonzero modes. Here we consider
Neumann and Dirichlet directions separately since there are
several crucial differences.

For the zero mode excitations along the Neumann directions, we
have the following correspondence\footnote{To be rigorous, the
directions $x^0,x^1,x^2,x^9$ are related to the original
coordinates by a conformal transformation after wick rotation as in the
radial quantization\cite{BMN}. This transformation leaves the 9 direction orthogonal
to the defect.}:
 \bea
{a^\dagger}_0^1 |0,p^+\rangle_{l.c.}  &\longleftrightarrow&
{1\over \sqrt{J}} \sum_{l=0}^{J}\frac{1}{N^{J/2+1}} \sigma^Z_{mn}
 \bar{q}^m Z^l (D_0 Z) Z^{J-l} q^n , \\
{a^\dagger}_0^2 |0,p^+\rangle_{l.c.}  &\longleftrightarrow&
{1\over \sqrt{J}} \sum_{l=0}^{J}\frac{1}{N^{J/2+1}} \sigma^Z_{mn} \bar{q}^m Z^l (D_1 Z) Z^{J-l} q^n, \\
{a^\dagger}_0^3 |0,p^+\rangle_{l.c.}  &\longleftrightarrow&
 {1\over \sqrt{J}} \sum_{l=0}^{J}\frac{1}{N^{J/2+1}} \sigma^Z_{mn} \bar{q}^m Z^l (D_2 Z) Z^{J-l} q^n, \\
\label{X^5}
  {a^\dagger}_0^5 |0,p^+\rangle_{l.c.}  &\longleftrightarrow&
  {1\over \sqrt{J}} \sum_{l=0}^{J}\frac{1}{N^{J/2+1}}
  \sigma^Z_{mn} \bar{q}^m Z^l X^5 Z^{J-l} q^n.
 \eea
The above open string states are associated with preserved
symmetries of the D5 brane. They are massive however since the
symmetries do not commute with the light cone Hamiltonian. Hence,
these operators are obtained from the vacuum operator
(\ref{vacuum}) by acting corresponding preserved symmetries in the
defect conformal field theory\cite{BMN,rey}. For example, the fourth operator
(\ref{X^5}) is obtained by acting a $SU(2)_H$ rotation on the
vacuum operator. The rotation also acts on the boundary fields
$\bar{q^m}$ and $q^n$ giving rise to additional terms such as \beq
\tilde{\sigma}^5_{mn}\bar{q}^m Z^{J+1} q^n \,. \eeq For notational
simplicity, we have suppressed this term in the above list.
Likewise, the other three operators have additional boundary
contributions. In the weak 't Hooft coupling regime, these
operators have $\Delta - J = 1$. Since they are in the same
multiplet as the chiral primary operator (\ref{vacuum}), their
dimensions are also protected by supersymmetry.

For nonzero mode excitations along the Neumann directions, we
insert operators with {\it cosine} phases\footnote{This point is
also noticed in \cite{BMN2}.}\footnote{In principle, we should assign phases including
the boundary contributions. Again, for simplicity, we suppress them since it does not affect
following conclusions.}
 \bea
{a^\dagger}_n^1 |0,p^+\rangle_{l.c.}  &\longleftrightarrow&
{1\over \sqrt{J}} \sum_{l=0}^{J}\frac{\sqrt{2}\cos\left({\pi nl \over J}\right)}{N^{J/2+1}} \sigma^Z_{mn} \bar{q}^m Z^l (D_0 Z) Z^{J-l} q^n,  \\
{a^\dagger}_n^2 |0,p^+\rangle_{l.c.}  &\longleftrightarrow&
{1\over \sqrt{J}} \sum_{l=0}^{J}\frac{\sqrt{2}\cos\left({\pi nl \over J}\right)}{N^{J/2+1}} \sigma^Z_{mn} \bar{q}^m Z^l (D_1 Z) Z^{J-l} q^n, \\
{a^\dagger}_n^3 |0,p^+\rangle_{l.c.}  &\longleftrightarrow&
{1\over \sqrt{J}} \sum_{l=0}^{J}\frac{\sqrt{2}\cos\left({\pi nl \over J}\right)}{N^{J/2+1}} \sigma^Z_{mn} \bar{q}^m Z^l (D_2 Z) Z^{J-l} q^n, \\
{a^\dagger}_n^5 |0,p^+\rangle_{l.c.}  &\longleftrightarrow&
{1\over \sqrt{J}} \sum_{l=0}^{J}\frac{\sqrt{2}\cos\left({\pi
nl \over J}\right)}{N^{J/2+1}} \sigma^Z_{mn} \bar{q}^m Z^l
X^5 Z^{J-l} q^n.
 \eea
The factor of $\sqrt{2}$ is necessary for correct normalization of
the free Feynman diagram in the two-point function. Notice that
unlike the closed string case, the operators with single
insertions are not trivially zero which reflects the fact that
there is no level matching condition for open strings.  In
addition, the sign of $n$ has no significance, which corresponds
to the fact that there is only one set of oscillators instead of
both the left and
right moving sectors.\\
We can compute the anomalous dimension of
these operators following the closed string case discussed in the
appendix of \cite{BMN}. The only difference from the closed string case
is that the exponential phase has been replaced by the cosine
phase. For example, let $\mathcal O$ be the fourth operator (\ref{X^5}) above.
The contribution from ${1\over 2\pi g} \int d^4x
2Tr[X^5ZX^5\bar{Z}]$ in the bulk action gives
 \bea \nonumber
 \langle {\mathcal O}(x){\mathcal O}^*(0) \rangle &=& \frac{\mathcal N}{|x|^{2\Delta}}
  \left[ 1+ {1\over J}\sum_{l=0}^{J-1} N(2\pi g)8\cos\left({\pi nl \over J}\right)\cos\left({\pi n(l+1)
  \over J}\right) {1\over 4\pi^2}\log|x|\Lambda \right] \\
  \nonumber
&=& \frac{\mathcal N}{|x|^{2\Delta}} \left[ 1+ {1\over
J}\sum_{l=0}^{J-1} N(2\pi g)4 \left\{ \cos\left({\pi n(2l+1) \over
J}\right)+\cos\left({\pi n \over J} \right)\right\} {1\over
4\pi^2}\log|x|\Lambda \right] \\  &=& \frac{\mathcal
N}{|x|^{2\Delta}} \left[ 1+ N(2\pi g)4\cos\left({\pi n \over J}
\right) {1\over 4\pi^2}\log|x|\Lambda \right],
 \eea
where $\mathcal N$ is a normalization factor and $\Lambda$ is the
UV cutoff scale. As argued in \cite{BMN}, contributions from other
Feynman diagrams cancel this contribution when $n=0$.  Therefore,
the full contribution can be taken into account by simply
replacing $\cos\left({\pi n \over J}\right)$ with
$\cos\left({\pi n \over J}\right)-1$. Finally, we have to the
leading order
 \beq
\langle{\mathcal O}(x){\mathcal O}^*(0) \rangle = \frac{\mathcal
N}{|x|^{2\Delta}} \left[ 1- \frac{\pi g N n^2}{J^2} \log|x|\Lambda
\right].
 \eeq
Therefore, one gets
 \beq (\Delta - J)_n = 1 + { \pi g N n^2
 \over 2J^2} = 1+ { n^2 \over 8 (\alpha^\prime p^+)^2}\,.
 \eeq
This is exactly the first order expansion of light-cone energy of
corresponding string states.

Now consider the directions with Dirichlet boundary conditions. As
mentioned earlier, the associated mode expansions do not have zero
modes.  For nonzero mode excitations, we insert appropriate
operators with {\it sine} phases as follows
 \bea
{a^\dagger}_n^4 |0,p^+\rangle_{l.c.}  &\longleftrightarrow&
{1\over \sqrt{J}} \sum_{l=0}^{J}\frac{\sqrt{2}\sin\left({\pi n l \over J}\right)}{N^{J/2+1}} \sigma^Z_{mn} \bar{q}^m Z^l (D_9 Z) Z^{J-l} q^n,  \\
{a^\dagger}_n^6 |0,p^+\rangle_{l.c.}  &\longleftrightarrow&
{1\over \sqrt{J}} \sum_{l=0}^{J}\frac{\sqrt{2}\sin\left({\pi n l \over J}\right)}{N^{J/2+1}} \sigma^Z_{mn} \bar{q}^m Z^l X^6 Z^{J-l} q^n, \\
{a^\dagger}_n^7 |0,p^+\rangle_{l.c.}  &\longleftrightarrow&
{1\over \sqrt{J}} \sum_{l=0}^{J}\frac{\sqrt{2}\sin\left({\pi n l \over J}\right)}{N^{J/2+1}} \sigma^Z_{mn} \bar{q}^m Z^l X^7 Z^{J-l} q^n, \\
{a^\dagger}_n^8 |0,p^+\rangle_{l.c.}  &\longleftrightarrow&
{1\over \sqrt{J}} \sum_{l=0}^{J}\frac{\sqrt{2}\sin\left({\pi n
l \over J}\right)}{N^{J/2+1}} \sigma^Z_{mn} \bar{q}^m Z^l
X^8 Z^{J-l} q^n.
 \eea
Notice that the sine phases naturally kill the zero modes when
$n=0$. We should ask what is the fate of the operators with
insertions along the Dirichlet directions without phase. These
operators are obtained by acting on the vacuum operator (\ref{vacuum}) with
symmetries broken by the defect\footnote{As a result, they do not
act on $q$ and $\bar{q}$ unlike the case for Neumann
directions.}. Therefore, their dimensions are not generally
protected. In fact, the operators $X^6,X^7,X^8$ are in the bulk
vector multiplet and couple to the defect hyper multiplet via 
defect F-term. Similarly, the normal
derivative $D_9 Z$ couples to the defect hyper multiplet despite
the fact that $Z$ itself is in the bulk hyper
multiplet\cite{ooguri1}. This boundary interaction gives rise to
large anomalous dimensions of order $N/J \sim J$ when one inserts
operators {\it without phases}. Hence such operators will
disappear in the strong 't Hooft coupling regime as implied by the
open string spectrum. Nevertheless, once we include the sine
phase, boundary interactions are suppressed by a factor of $\sin^2
\left({\pi n \over J}\right) \sim 1/J^2$, and they can be ignored
to the leading order in $1/J$. Therefore, the only contribution to
anomalous dimensions comes from the bulk interaction. The
computation is essentially the same as above, and the result
agrees with the open string spectrum.

For fermionic excitations, we insert $J=1/2$ components of
$\lambda^{im}$, which is just $\lambda^{i1}. $\footnote{We take
$m$ to be the quantum number of $J$, which is a generator of
Cartan subalgebra of $SU(2)_H$.} As in the bosonic sector, the
number of zero modes is half of that of non-zero modes. Hence, we
need a similar mechanism to remove possible gauge theory operators
corresponding to the 4 unphysical zero modes. The symmetry
breaking pattern and the form of defect couplings in the
defect CFT again allow one to do this consistently. Recall that the
operators $\lambda^{i1}_1$ and $\lambda^{i1}_2$ are in the vector
and hyper multiplets respectively. Only $\lambda^{i1}_1$ couples
to the defect hypermultiplet while $\lambda^{i1}_2$ can be
obtained from $Z$ by acting preserved supersymmetries 
\footnote{They also transform $q$ and $\bar{q}$ into $\Psi$ and
$\bar{\Psi}$. Therefore, when we insert $\lambda^{i1}_2$, we have
additional boundary terms with $q$ or $\bar{q}$ replaced by $\Psi$
or $\bar{\Psi}$.}. Therefore, we associate sine and cosine phases
with $\lambda^{i1}_1$ and $\lambda^{i1}_2$ respectively. As in the
bosonic sector, this assignment reproduces the open string
spectrum in the fermionic sector.

\section{Conclusion}
In this article, we have considered a Penrose limit of type IIB
string theory on $AdS_5 \times S^5$ with a D5-brane spanning an 
$AdS^4 \times S^2$ whose dual field theory is ${\mathcal N}=4$ SYM
coupled to a 3-dimensional conformal defect. The Penrose limit
gives rise to a D5-brane in the pp-wave background. The limit
corresponds to looking at a subsector of operators in the dual
field theory with large $J \sim \sqrt{N}$ and finite $\Delta-J$ in
the large 't Hooft coupling regime. We have studied perturbative
open string spectrum on this brane and constructed a complete list
of gauge invariant operators dual to the open string states from
the defect conformal field theory. The peculiar feature of defect
couplings, symmetry breaking pattern in the dual field theory, and
sine-cosine phases are essential to reproduce the proper boundary
conditions for the open strings.

One can also consider several M D5-branes. Then the defect
hypermultiplet gets an additional $U(M)$ index with $q^m$ and
$\bar{q}^n$ transforming as {\bf M} and {\bf \=M} of $U(M)$
respectively. This naturally induces Chan-Paton factors at the
ends of open strings as expected.

It would be interesting to construct defect conformal field
theories arising from other supersymmetric brane intersections and
study their Penrose limits.  Then we expect to find specific
defect couplings and symmetry breaking patterns which reflect the boundary
conditions of the D-branes in this limit.

\begin{center}
\bf{Acknowledgments}
\end{center}
We would like to thank David Berenstein, Jaume Gomis, Sangmin Lee, Yuji Okawa, 
Hirosi Ooguri, Harlan Robins, John Schwarz, and Soojin Son for useful discussions and comments. This research was supported in part by DOE grant
DE-FG03-92-ER40701.

\bibliography{ppwave4}
\bibliographystyle{ssg}

\end{document}